\documentclass[a4paper,10 pt,conference,dvipsnames]{ieeeconf}

\IEEEoverridecommandlockouts

\usepackage{geometry}
\renewcommand{\baselinestretch}{0.995}
\usepackage{mathptmx}
\usepackage{times}
\usepackage{textpos}

\usepackage[usenames,dvipsnames]{xcolor}

\usepackage{amsmath,amssymb,mathtools}

\usepackage{amsthm}
\usepackage{bbold,dsfont,bbm}

\usepackage{graphicx}
\usepackage{tikz,tikz-cd}

\usepackage{hyperref}
\usepackage[capitalise,noabbrev]{cleveref}
\usepackage{cite}

\usepackage{multirow}
\usepackage{booktabs}

\usepackage[acronym]{glossaries}

\usepackage{comment}

\usepackage{units}

\usepackage[font=footnotesize]{caption}
\usepackage{subcaption}

\newcommand{\cG}{\mathcal{G}}

\newcommand{\cM}{\mathcal{M}}

\newcommand{\cV}{\mathcal{V}}

\definecolor{lightblue}{rgb}{0.60784,0.76078,0.90196}
\definecolor{darkblue}{rgb}{0.26667,0.44706,0.76863}
\definecolor{lightgreen}{rgb}{0.66275,0.81569,0.55686}
\definecolor{darkgreen}{rgb}{0.43922,0.67843,0.27843}
\definecolor{orange}{rgb}{0.92941,0.49020,0.19216}
\definecolor{yellow}{rgb}{1.00000,0.75294,0.00000}
\definecolor{grey}{rgb}{0.64706,0.64706,0.64706}
\definecolor{purple}{rgb}{0.51373,0.23529,0.04706}
\newacronym{abk:amod}{AMoD}{Autonomous Mobility-on-Demand}
\newacronym{abk:iamod}{\mbox{I-AMoD}}{intermodal \gls{abk:amod}}
\newacronym{abk:av}{\mbox{AV}}{autonomous vehicle}

\newacronym{abk:bpr}{BPR}{Bureau of Public Roads}
\newacronym{abk:bev}{BEV}{Battery Electric Vehicle}
\newacronym{abk:ca}{CA}{congestion-aware}
\newacronym{abk:cara}{CARS}{congestion-aware routing scheme}
\newacronym{abk:cpo}{CPO}{complete partial order}
\newacronym{abk:cdp}{CDP}{co-design problem}
\newacronym{abk:cdpi}{CDPI}{co-design problem with implementation}
\newacronym{abk:dp}{DP}{design problem}
\newacronym{abk:dpi}{DPI}{design problem with implementation}
\newacronym{abk:dcpo}{DCPO}{directed complete partial order}
\newacronym{abk:es}{ES}{e-scooter}
\newacronym{abk:ffcs}{FFCS}{free floating car sharing systems}
\newacronym{abk:ghg}{GHG}{greenhouse gas}
\newacronym{abk:icev}{ICEV}{ Internal Combustion Engine Vehicle}
\newacronym{abk:kpi}{KPIs}{Key Performance Indicators}
\newacronym{abk:lw}{LW}{Lightweight}
\newacronym{abk:mm}{{$\mu$}M}{micromobility}
\newacronym{abk:mod}{MoD}{Mobility-on-Demand}
\newacronym{abk:mcdp}{MCDP}{Monotone Co-Design Problem}
\newacronym{abk:mcfp}{MCFP}{multi-commodity flow problem}
\newacronym{abk:nyc}{NYC}{New York City}
\newacronym{abk:poset}{poset}{partially ordered set}
\newacronym{abk:sb}{SB}{shared bike}
\newacronym{abk:spp}{SPP}{shortest path problem}
\newacronym{abk:kdspp}{k-dSPP}{k-disjoint \gls{abk:spp}}
\newacronym{abk:su}{SU}{Sport Utility}

\newcommand{\achievableSpeedVeh}{v_\mathrm{V,a}}

\newcommand{\arcBaselineUsage}{u_{ij}}
\newcommand{\arcEnergy}{e_{ij}}
\newcommand{\arcNominalCapacity}{c_{ij}}
\newcommand{\arcLength}{s_{ij}}
\newcommand{\arcSpeedVeh}{v_{\mathrm{V},ij}}

\newcommand{\arcSpeedLimitVeh}{v_{\mathrm{L,V},ij}}

\newcommand{\arcTime}{t_{ij}}
\newcommand{\averageTravelTime}{t_\mathrm{avg}}

\newcommand{\costFixVeh}{C_\mathrm{V,f}}

\newcommand{\costFixTrain}{C_\mathrm{S,f}}
\newcommand{\costOpTrain}{C_\mathrm{S,o}}
\newcommand{\costOpVeh}{C_\mathrm{V,o}}

\newcommand{\costTot}{C_\mathrm{tot}}
\newcommand{\costVeh}{C_\mathrm{V}}
\newcommand{\costSub}{C_\mathrm{S}}

\newcommand{\distanceVeh}{s_\mathrm{V,tot}}

\newcommand{\emissionsVeh}{m_\mathrm{CO_2,V,tot}}

\newcommand{\emissionsTot}{m_\mathrm{CO_2,tot}}
\newcommand{\emissionsSub}{m_\mathrm{CO_2,S}}
\newcommand{\exec}[1]{\textsf{\textup{exe}}_{#1}}
\newcommand{\eval}[1]{\textsf{\textup{eva}}_{#1}}
\newcommand{\flow}[2]{f_m\left(#1,#2\right)}
\newcommand{\flowTot}[2]{f_\mathrm{tot}\left(#1,#2\right)}

\newcommand{\flowReba}[2]{f_{0}\left(#1,#2\right)}

\newcommand{\freqTrain}{\varphi_j}
\newcommand{\freqTrainBaseline}{\varphi_{j,\mathrm{base}}}
\newcommand{\functionality}[1]{\textsf{\textup{f}}_{#1}}
\newcommand{\maph}[1]{h_{#1}}
\newcommand{\implementation}[1]{\textsf{\textup{i}}_{#1}}
\newcommand{\lifeVeh}{l_\mathrm{V}}
\newcommand{\lifeTrain}{l_\mathrm{S}}

\newcommand{\numberFleetVeh}{n_\mathrm{V,max}}

\newcommand{\numberFleetUsedVeh}{n_\mathrm{V,u}}

\newcommand{\numberFleetTrain}{n_\mathrm{S}}
\newcommand{\numberFleetTrainBaseline}{n_\mathrm{S,base}}

\newcommand{\rplusbar}{\overline{\mathbb{R}}_+}
\newcommand{\nbar}{\overline{\mathbb{N}}}

\newcommand{\traveltime}{t}

\newcommand{\timePedestrianSubway}{\traveltime_{\mathrm{WS}}}
\newcommand{\timeSubwayPedestrian}{\traveltime_{\mathrm{SW}}}

\newcommand{\timeRoadPedestrian}{\traveltime_{\mathrm{RW}}}

\newcommand{\timePedestrianRoad}{\traveltime_{\mathrm{WR}}}

\newcommand{\setOfArcs}{\mathcal{A}}

\newcommand{\setOfArcsRoad}{\mathcal{A}_{\mathrm{R}}}
\newcommand{\setOfArcsSubway}{\mathcal{A}_{\mathrm{P}}}
\newcommand{\setOfArcsPedestrian}{\mathcal{A}_{\mathrm{W}}}
\newcommand{\setOfArcsCommute}{\mathcal{A}_{\mathrm{C}}}

\newcommand{\setOfFunctionalities}[1]{\mathcal{F}_{#1}}

\newcommand{\GraphRoad}{\mathcal{G}_\mathrm{R}}
\newcommand{\GraphSubway}{\mathcal{G}_\mathrm{P}}
\newcommand{\GraphPedestrian}{\mathcal{G}_\mathrm{W}}

\newcommand{\setOfImplementations}[1]{\mathcal{I}_{#1}}
\newcommand{\setOfRequestsNumber}{\mathcal{M}}
\newcommand{\setOfResources}[1]{\mathcal{R}_{#1}}
\newcommand{\setOfVertices}{\mathcal{V}}

\newcommand{\setOfVerticesRoad}{\mathcal{V}_{\mathrm{R}}}

\newcommand{\setOfVerticesSubway}{\mathcal{V}_{\mathrm{P}}}
\newcommand{\setOfVerticesPedestrian}{\mathcal{V}_{\mathrm{W}}}
\newcommand{\arc}{(i,j)}

\newcommand{\bool}[1]{\mathds{1}_{#1}}

\newcommand{\true}[1]{\texttt{T}}
\newcommand{\setOfResourcesAntichain}[1]{\textsf{\textup{A}}\setOfResources{{#1}}}

\newtheorem{theorem}{Theorem}[section]

\newtheorem{lemma}[theorem]{Lemma}

\theoremstyle{definition}
\newtheorem{definition}{Definition}[section]
\theoremstyle{remark}
\newtheorem*{remark}{Remark}

\makeatletter
    \if@cref@capitalise
        \crefname{subsection}{Section}{Sections}
        \crefname{subsubsection}{Section}{Sections}
        \crefname{assump}{Assumption}{Assumptions}
        \crefname{problem}{Problem}{Problems}
    \else
        \crefname{subsection}{section}{sections}
        \crefname{subsubsection}{section}{sections}
        \crefname{assump}{assumption}{assumptions}
        \crefname{problem}{problem}{problems}
    \fi
\makeatother

\usetikzlibrary{positioning,intersections,hobby,patterns,calc,decorations.pathmorphing,decorations.markings, shadows,shapes}
\tikzstyle{block} = [draw, rectangle, minimum height=2em, minimum width=3em]
\tikzstyle{block1} = [draw, rectangle, minimum height=1.5em, minimum width=2.5em]
\tikzstyle{blockDyn} = [draw, rectangle, minimum height=2.5em, minimum width=3.5em, align=center, inner sep=10pt, thick, fill=white, copy shadow={draw=black,fill=black,opacity=1,shadow xshift=0.5ex,shadow yshift=-0.5ex}]
\tikzstyle{blockAlg} = [draw, rectangle, minimum height=1.5em, minimum width=2.5em, align=center, inner sep=10pt, thick]
\tikzstyle{sum} = [draw,circle]
\tikzstyle{nodePre} = [circle, draw,inner sep=1pt,node contents={$\preceq$}]
\tikzstyle{nodePreEmpty} = [circle, draw,inner sep=1pt]
\tikzstyle{nodePos} = [circle, draw,inner sep=1pt,node contents={$\posceq$}]
\tikzstyle{nodeProd} = [rectangle, draw,inner sep=4pt,node contents={$\times$}]
\tikzstyle{nodeSum} = [rectangle, draw,inner sep=4pt,node contents={$+$}]

\definecolor{DPgreen}{RGB}{34,139,34}

\newif\ifmargincomments %

\newif\ifextendedversion %

\ifmargincomments

\else

\fi

\title{
\textbf{On the Co-Design of AV-Enabled Mobility Systems}
}
\author{Gioele Zardini$^{1,3}$, Nicolas Lanzetti$^{2,3}$, Mauro Salazar$^{3,4}$, Andrea Censi$^1$,
	Emilio Frazzoli$^{1}$,  and Marco Pavone$^3$
\thanks{$^1$Institute for Dynamic Systems and Control,
        ETH Z\"urich,
        {\tt \{gzardini,acensi,emilio.frazzoli\}@ethz.ch}}%
\thanks{$^2$Automatic Control Laboratory,
        ETH Z\"urich,
        {\tt lnicolas@ethz.ch}}%
\thanks{$^3$Department of Aeronautics and Astronautics, Stanford University,
        {\tt pavone@stanford.edu}}
\thanks{$^4$Control Systems Technology Group, Eindhoven University of Technology, {\tt m.r.u.salazar@tue.nl}}
\thanks{A preliminary version of this paper was presented at the 99th Annual Meeting of the Transportation Research Board~\cite{ZardiniLanzettiEtAl2020}.}
\thanks{This research was supported by the National Science Foundation under CAREER Award CMMI-1454737, the Toyota Research Institute (TRI), and ETH Z\"urich. This article solely reflects the opinions and conclusions of its authors and not NSF, TRI, or any other entity.}}

\begin{document}

\maketitle
\begin{abstract}
The design of \glspl{abk:av} and the design of \gls{abk:av}-enabled mobility systems are closely coupled. Indeed, knowledge about the intended service of \glspl{abk:av} would impact their design and deployment process, whilst insights about their technological development could significantly affect transportation management decisions.
This calls for tools to study such a coupling and co-design \glspl{abk:av} and \gls{abk:av}-enabled mobility systems in terms of different objectives.
In this paper, we instantiate a framework to address such co-design problems.
In particular, we leverage the recently developed theory of co-design to frame and solve the problem of designing and deploying an intermodal Autonomous Mobility-on-Demand system, whereby \glspl{abk:av} service travel demands jointly with public transit, in terms of fleet sizing, vehicle autonomy, and public transit service frequency. Our framework is modular and compositional, allowing one to describe the design problem as the interconnection of its individual components and to tackle it from a system-level perspective.
To showcase our methodology, we present a real-world case study for Washington D.C., USA.
Our work suggests that it is possible to create user-friendly optimization tools to systematically assess costs and benefits of interventions, and that such analytical techniques might gain a momentous role in policy-making in the future.
\end{abstract}

\section{Introduction}
\label{sec:introduction}
Arguably, the current design process for \glspl{abk:av} largely suffers from the lack of clear, specific requirements in terms of the service such vehicles will be providing.  Yet, knowledge about their intended service (e.g., last-mile versus point-to-point travel) might dramatically impact how the AVs are designed, and, critically, significantly ease their development process. For example, if for a given city we knew that for an effective on-demand mobility system autonomous cars only need to drive up to \unit[25]{mph} and only on relatively easy roads, their design would be greatly simplified and their deployment could certainly be accelerated.
At the same time, from the system-level perspective of transportation management, knowledge about the trajectory of technology development for  \glspl{abk:av} would certainly impact decisions on infrastructure investments and provision of service. 
In other words, the design of the \glspl{abk:av} and the design of a mobility system leveraging \glspl{abk:av} are intimately {\em coupled}. 
This calls for methods  to reason about such a coupling, and in particular to \emph{co-design} the \glspl{abk:av} and the associated \gls{abk:av}-enabled mobility system. A key requirement in this context is the ability to account for a range of heterogeneous objectives that are often not directly comparable (consider, for instance, travel time and emissions). 

Accordingly, the goal of this paper is to lay the foundations for a framework through which one can co-design future \gls{abk:av}-enabled mobility systems. 
Specifically, we show how one can leverage the recently developed mathematical theory of co-design~\cite{Censi2015,Censi2016,Censi2017b}, which provides a general methodology to co-design complex systems in a modular and compositional fashion.
This tool delivers the set of rational design solutions lying on the Pareto front, allowing one to reason about costs and benefits of the individual design options.
The framework is instantiated in the setting of co-designing intermodal  \gls{abk:amod} systems~\cite{SalazarLanzettiEtAl2019}, whereby fleets of self-driving vehicles provide on-demand mobility jointly with public transit. Aspects subject to co-design include fleet size, \gls{abk:av}-specific characteristics, and public transit service frequency.

\subsection{Literature Review}
Our work lies at the interface of the design of urban public transportation services and the design of \gls{abk:amod} systems. The first research stream is reviewed in~\cite{FarahaniMiandoabchiEtAl2013,GuihaireHao2008}, and comprises \emph{strategic} long-term infrastructure modifications and \emph{operational} short-term scheduling. The joint design of traffic network topology and control infrastructure has been presented in~\cite{CongDeSchutterEtAl2015}.
Public transportation scheduling has been solved jointly with the design of the transit network in a passengers' and operators' cost-optimal fashion in~\cite{Arbex2015}, using demand-driven approaches in~\cite{Sun2014}, and in an energy-efficient way in~\cite{Su2013}.
However, these works only focus on the public transit system and do not consider its joint design with an \gls{abk:amod} system.
The research on the design of \gls{abk:amod} systems is reviewed in~\cite{NarayananEtAl2020} and mainly pertains their fleet sizing.
In this regard, studies range from simulation-based approaches~\cite{BarriosGodier2014,FagnantKockelman2018,Vazifeh2018,Boesch2016} to analytical methods~\cite{SpieserTreleavenEtAl2014}.
 In \cite{ZhangSheppardEtAl2018}, the authors jointly design the fleet size and the charging infrastructure, and formulate the arising design problem as a mixed integer linear program.
 The authors of \cite{BeaujonTurnquist1991} solve the fleet sizing problem together with the vehicle allocation problem.
 Finally, \cite{Wallar2019} co-designs the \gls{abk:amod} fleet size and its composition. 
 More recently, the joint design of multimodal transit networks and \gls{abk:amod} systems was formulated in~\cite{PintoHylandEtAl2019} as a bilevel optimization problem and solved with heuristics.
Overall, the problem-specific structure of existing design methods for \gls{abk:amod} systems is not amenable to a modular and compositional problem formulation. Moreover, previous work does not capture important aspects of \gls{abk:av}-enabled mobility systems, such as other transportation modes and \gls{abk:av}-specific design parameters (e.g., the level of autonomy). 

\subsection{Statement of Contribution}
In this paper we lay the foundations for the systematic study of the design of \gls{abk:av}-enabled mobility systems. Specifically, we leverage the mathematical theory of co-design~\cite{Censi2015} to devise a framework to study the design of \gls{abk:iamod} systems in terms of fleet characteristics and public transit service, enabling the computation of the \emph{rational} solutions lying on the Pareto front of minimal travel time, transportation costs, and emissions.
Our framework allows one to structure the design problem in a modular way, in which each different transportation option can be ``plugged in'' in a larger model. 
Each model has minimal assumptions: Rather than properties such as linearity and convexity, we ask for very general monotonicity assumptions. For example, we assume that the cost of automation increases monotonically with the speed achievable by the \gls{abk:av}.
We are able to obtain the full Pareto front of \emph{rational} solutions, or, given policies, to weigh incomparable costs (such as travel time and emissions) and to present actionable information to the stakeholders of the mobility ecosystem.
We showcase our methodology through a real-world case study of Washington D.C., USA. We show how, given the model, we can easily formulate and answer several questions regarding the introduction of new technologies and investigate possible infrastructure interventions.
\subsection{Organization}
The remainder of this paper is structured as follows: \cref{sec:background} reviews the mathematical theory of co-design.
\cref{sec:codesignav} presents the co-design problem for \gls{abk:av}-enabled mobility systems.
We showcase our approach with real-world case studies for Washington D.C., USA, in \cref{sec:results}. \cref{sec:conclusion} concludes the paper with a discussion and an overview on future research directions.
\section{Background}
\label{sec:background}
This paper builds on the mathematical theory of co-design, presented in~\cite{Censi2015}. In this section, we present a review of the main contents needed for this work. 

\subsection{Orders}
We will use basic facts from order theory, which we review in the following.
\begin{definition}[Poset]
A \gls{abk:poset} is a tuple $\langle \mathcal{P},\preceq_\mathcal{P} \rangle$, where $\mathcal{P}$ is a set and 	$\preceq_\mathcal{P}$ is a partial order, defined as a reflexive, transitive, and antisymmetric relation.
\end{definition}
Given a \gls{abk:poset}, we can formalize the idea of ``Pareto front'' through antichains.
\begin{definition}[Antichains]
A subset $S\subseteq\mathcal{P}$ is an antichain iff
no elements are comparable: For $x,y\in S$, $x\preceq y$ implies $x=y$.	We denote by $\textsf{A}\mathcal{P}$ the set of all antichains in $\mathcal{P}$.
\end{definition}
\begin{definition}[Directed set]
A subset $S\subseteq\mathcal{P}$ is directed if each pair of elements in $S$ has an upper bound: For all $a,b\in S$, there exists a $c\in S$ such that $a\preceq c$ and $b\preceq c$.
\end{definition}
\begin{definition}[Completeness]
	A \gls{abk:poset} is a \gls{abk:cpo} if each of its directed subsets has a supremum and a least element.
\end{definition}
For instance, the poset $\langle \mathbb{R}_+,\leq\rangle$, with $\mathbb{R}_+\coloneqq\{x\in\mathbb{R}\,|\,x\geq 0\}$, is not complete, as its directed  subset $\mathbb{R}_+\subseteq\mathbb{R}_+$ does not have an upper bound (and therefore a supremum). Nonetheless, we can make it complete by artificially adding a top element $\top$, i.e., by defining $\langle\rplusbar,\leq\rangle$ with $\rplusbar\coloneqq\mathbb{R}_+\cup \{\top\}$ and $a\leq \top$ for all $a\in\mathbb{R}_+$. Similarly, we can complete $\mathbb{N}$ to $\nbar$.

In this setting, Scott-continuous maps will play a key role. Intuitively, Scott-continuity can be understood as a stronger notion of monotonicity. 
%
\begin{definition}[Scott continuity]
A map $f:\mathcal{P}\rightarrow \mathcal{Q}$ between two posets $\langle \mathcal{P}, \preceq_\mathcal{P} \rangle$ and $\langle \mathcal{Q}, \preceq_\mathcal{Q} \rangle$ is Scott-continuous iff for each directed set $D\subseteq\mathcal{P}$ the image $f(D)$ is directed and $\sup f(D)=f(\sup D)$.
\end{definition}

\subsection{Mathematical Theory of Co-Design}
We start by presenting design problems with implementation (\glsunset{abk:dpi}\glspl{abk:dpi}), which can then be composed and interconnected to form a  \gls{abk:cdpi}. 
\begin{definition}[\gls{abk:dpi}]
A \gls{abk:dpi} is a tuple $\langle \setOfFunctionalities{},\setOfResources{},\setOfImplementations{},\exec{},\eval{}\rangle$:
\begin{itemize}
	\item $\setOfFunctionalities{}$ is a poset, called functionality space;
	\item $\setOfResources{}$ is a poset, called resource space;
	\item $\setOfImplementations{}$ is a set, called implementation space;
	\item the map $\exec{}:\setOfImplementations{}\to\setOfFunctionalities{}$ maps
an implementation to the functionality it provides;
	\item the map $\eval{}:\setOfImplementations{}\to\setOfResources{}$, maps an implementation to the resources it requires.
\end{itemize}
\end{definition}
Given a \gls{abk:dpi} we can define a map which, given a functionality $\functionality{}\in\setOfFunctionalities{}$, returns all the non-comparable resources (i.e., the antichain) which provide $\functionality{}$.
\begin{definition}[Functionality to resources map]
\label{def:map h}
	Given a \gls{abk:dpi} $\langle \setOfFunctionalities{},\setOfResources{},\setOfImplementations{},\exec{},\eval{} \rangle$ define the map $\maph{}:\setOfFunctionalities{}\to\setOfResourcesAntichain{}$ as
	\begin{equation}\label{eq:map h}
		\begin{aligned}
		\maph{}:\:
		&\setOfFunctionalities{} &\to && &\setOfResourcesAntichain{} \\
		&\functionality{} &\mapsto && &\min_{\preceq_{\setOfResources{}}}\{\eval{}(\implementation{})\,|\, \implementation{}\in\setOfImplementations{}\wedge \functionality{}\preceq\exec{}(\implementation{})\}.
		\end{aligned}
	\end{equation}
\end{definition}
In particular, if a functionality is infeasible, then $\maph{}(\functionality{})=\emptyset$. We now turn our attention to ``monotone'' \glspl{abk:dpi}.
\begin{definition}[Monotone \gls{abk:dpi}]
We say a \gls{abk:dpi} $\langle \setOfFunctionalities{},\setOfResources{},\setOfImplementations{},\exec{},\eval{} \rangle$ is monotone if:
\begin{enumerate}
	\item The posets $\setOfFunctionalities{}$ and $\setOfResources{}$ are \glspl{abk:cpo}.
	\item The map $\maph{}$ (see \cref{def:map h}) is Scott-continuous.
\end{enumerate}
\end{definition}
Individual \glspl{abk:dpi} can be composed in series (i.e., the functionality of a \gls{abk:dpi} is the resource of a second \gls{abk:dpi}) and in parallel (i.e., two \glspl{abk:dpi} share the same resource or functionality) to obtain a \gls{abk:cdpi}. Notably, such compositions preserve monotonicity and, thus, all related algorithmic properties. For further details we refer to \cite{Censi2015}.

\section{Co-Design of AV-enabled Mobility Systems}
\label{sec:codesignav}
\subsection{Intermodal AMoD Framework}
\label{subsec:iamod}
\subsubsection{Multi-Commodity Flow Model}
\begin{figure}[tb]
	\begin{center}
		\includegraphics[width=0.85\columnwidth]{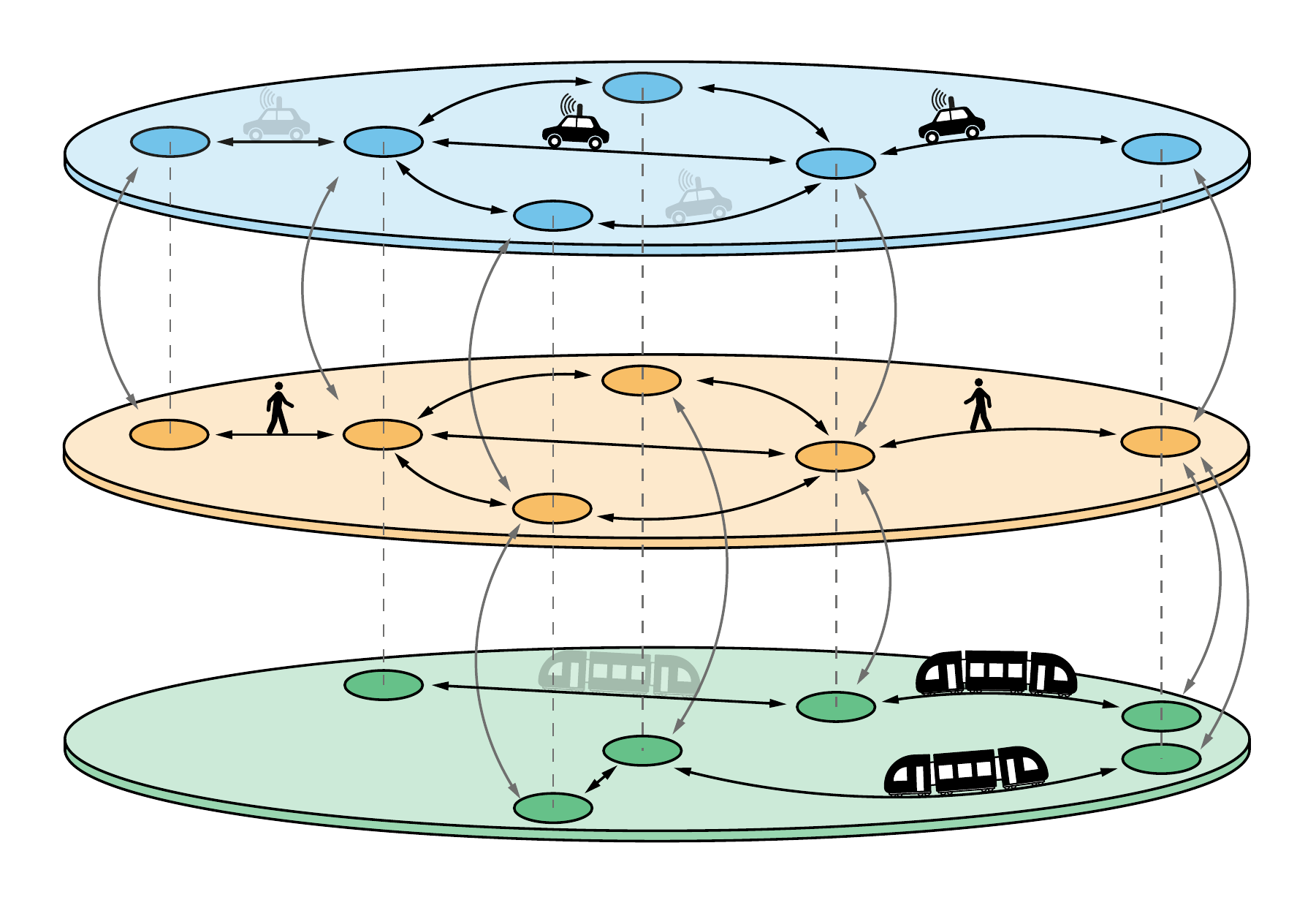}
		\caption{The \gls{abk:iamod} network consists of a road, a walking, and a public transportation digraph. The coloured circles represent stops or intersections and the black arrows denote road links, pedestrian pathways, or public transit arcs. Dashed lines are nodes which are close geographically, while grey arrows denote the mode-switching arcs connecting them\protect\footnotemark.}
		\label{fig:iamod}
		\vspace{-6mm}
	\end{center}
\end{figure}

The transportation system and its different modes are modeled using the digraph $\cG=\left( \setOfVertices, \setOfArcs \right)$, shown in \cref{fig:iamod}.
\footnotetext{We thank Ms. Sonia Monti for the illustration.}
It is described through a set of nodes $\setOfVertices$ and a set of arcs $\setOfArcs \subseteq \setOfVertices \times \setOfVertices$.
Specifically, it contains a road network layer $\GraphRoad=\left( \setOfVerticesRoad, \setOfArcsRoad\right)$, a public transportation layer $\GraphSubway=\left( \setOfVerticesSubway, \setOfArcsSubway\right)$, and a walking layer $\GraphPedestrian=\left( \setOfVerticesPedestrian, \setOfArcsPedestrian\right)$.
The road network is characterized through intersections $i \in \setOfVerticesRoad$ and road segments $(i,j)\in \setOfArcsRoad$. Similarly, public transportation lines are modeled through station nodes $i\in \setOfVerticesSubway$ and line segments $\arc \in \setOfArcsSubway$.
The walking network contains walkable streets $\arc \in \setOfArcsPedestrian$, connecting intersections $i\in \setOfVerticesPedestrian$.
Our model allows mode-switching arcs $\setOfArcsCommute \subseteq \setOfVerticesRoad \times \setOfVerticesPedestrian \cup \setOfVerticesPedestrian \times \setOfVerticesRoad \cup \setOfVerticesSubway \times \setOfVerticesPedestrian \cup \setOfVerticesPedestrian \times \setOfVerticesSubway$, connecting the road and the public transportation layers to the walking layer.
Consequently, $\setOfVertices=\setOfVerticesPedestrian \cup \setOfVerticesRoad \cup  \setOfVerticesSubway$ and $\setOfArcs=\setOfArcsPedestrian \cup \setOfArcsRoad \cup \setOfArcsSubway \cup \setOfArcsCommute$. Consistently with the structural properties of road and walking networks in urban environments, we assume the graph $\cG$ to be strongly connected. We represent customer movements by means of travel requests. A travel request refers to a customer flow starting its trip at a node $o\in\setOfVertices$ and ending it at a node $d\in\setOfVertices$.
\begin{definition}[Travel request]
A travel request $\rho$ is a triple $(o,d,\alpha) \in \setOfVertices \times \setOfVertices \times \mathbb{R}_+$, described by an origin node $o\in \setOfVertices$, a destination node $d \in \setOfVertices$, and the request rate $\alpha >0$, namely, the number of customers who want to travel from $o$ to $d$ per unit time.
\end{definition}
To ensure that a customer is not forced to use a given transportation mode, we assume all requests to lie on the walking digraph, i.e.,  $o_m,d_m \in \setOfVerticesPedestrian $ for all $m \in \cM \coloneqq\{1,\ldots,M\}$.

The flow $\flow{i}{j}\geq 0$ represents the number of customers per unit time traversing arc $\arc \in \setOfArcs$ and satisfying a travel request $m$. Furthermore, $\flowReba{i}{j}\geq 0$ denotes the flow of empty \glspl{abk:av} on road arcs $\arc \in \setOfArcsRoad$. This accounts for rebalancing flows of \glspl{abk:av} between a customer's drop-off and the next customer's pick-up. Assuming \glspl{abk:av} to carry one customer at a time, the flows satisfy
\allowdisplaybreaks
\begin{subequations}
	\label{eq:flowconstotal}
	\label{eq:IAMoDConservationNonNeg}
	\begin{align}
		&\sum_{i:\arc\in\setOfArcs}\flow{i}{j} + \bool{j=o_m}\cdot \alpha_m = \sum_{k:(j,k)\in\setOfArcs}\flow{j}{k} + \bool{j=d_m}\cdot \alpha_m\span\span\nonumber\\
		&  \hspace{4.8cm} \forall m\in\cM,\, j\in\cV \label{eq:flowconsa}\\
		&\sum_{i:\arc\in\setOfArcsRoad}\flowTot{i}{j} = \sum_{k:(j,k)\in\setOfArcsRoad}\flowTot{j}{k} \quad \forall j\in\setOfVerticesRoad,\label{eq:flowconsb}
	\end{align}
\end{subequations}
where $\mathbb{1}_{j=x}$ denotes the boolean indicator function and $\flowTot{i}{j}\coloneqq\flowReba{i}{j} + \sum_{m\in\setOfRequestsNumber}\flow{i}{j}$. Specifically, \eqref{eq:flowconsa} guarantees flows conservation for every transportation demand, and \eqref{eq:flowconsb} preserves flow conservation for \glspl{abk:av} on every road node.
Combining conservation of customers~\eqref{eq:flowconsa} with the conservation of \glspl{abk:av}~\eqref{eq:flowconsb} guarantees rebalancing \glspl{abk:av} to match the demand.

\subsection{Travel Time and Travel Speed}
\label{subsec:traveltime}
The variable $\arcTime$ denotes the time needed to traverse an arc $\arc$ of length $\arcLength$. We assume a constant walking speed on walking arcs and infer travel times on public transportation arcs from the public transit schedules. Assuming that the public transportation system at node $j$ operates with frequency $\freqTrain$, switching from a pedestrian vertex $i \in \setOfVerticesPedestrian$ to a public transit station $j \in \setOfVerticesSubway$ takes, on average, 
\begin{equation}
	\arcTime=\timePedestrianSubway+0.5\cdot 1/\freqTrain \quad \forall \arc \in \setOfArcsPedestrian,
\end{equation}
where $\timePedestrianSubway$ is a constant sidewalk-to-station travel time. 
We assume that the average waiting time for \gls{abk:amod} vehicles is $\timePedestrianRoad$, and that switching from the road graph and the public transit graph to the walking graph takes the transfer times $\timeRoadPedestrian$ and $\timeSubwayPedestrian$, respectively. 
While each road arc $\arc \in \setOfArcsRoad$ is characterized by a speed limit $\arcSpeedLimitVeh$, \glspl{abk:av} safety protocols impose a maximum achievable velocity $\achievableSpeedVeh$. 
In order to prevent too slow and therefore dangerous driving behaviours, we only consider road arcs through which the \glspl{abk:av} can drive at least at a fraction $\beta$ of the speed limit: Arc $\arc \in \setOfArcsRoad$ is kept in the road network iff 
\begin{equation}
    \label{eq:droparcs}
    \achievableSpeedVeh \geq \beta \cdot \arcSpeedLimitVeh,
\end{equation}
where $\beta\in (0,1]$. We set the velocity of all arcs fulfilling condition~\eqref{eq:droparcs} to $\arcSpeedVeh=\min\{ \achievableSpeedVeh, \arcSpeedLimitVeh\}$ and compute the travel time to traverse them as
\begin{equation}
	\arcTime=\arcLength/\arcSpeedVeh \quad \forall \arc \in \setOfArcsRoad.
\end{equation}
\subsection{Road Congestion}
We capture congestion effects with a threshold model. The total flow on each road arc $\arc \in \setOfArcsRoad$, given by the sum of the \glspl{abk:av} flow $\flowTot{i}{j}$ and the baseline usage $\arcBaselineUsage$ (e.g., private vehicles), must remain below the nominal capacity $\arcNominalCapacity$ of the arc:
\begin{equation}
\label{eq:capacity}
	\flowTot{i}{j}+\arcBaselineUsage \leq \arcNominalCapacity \quad \forall \arc \in \setOfArcsRoad.
\end{equation}
\subsection{Energy Consumption}
We compute the energy consumption of \glspl{abk:av} for each road link considering an urban driving cycle, scaled so that the average speed $v_\mathrm{avg,cycle}$ matches the free-flow speed on the link. The energy consumption is then scaled as
\begin{equation}
\arcEnergy=e_\mathrm{cycle} \cdot \arcLength/s_\mathrm{cycle} \quad \forall \arc \in \setOfArcsRoad.
\end{equation}
For the public transportation system, we assume a constant energy consumption per unit time. This approximation is reasonable in urban environments, as the operation of the public transportation system is independent from the number of customers serviced, and its energy consumption is therefore customer-invariant.
\subsection{Fleet Size}
We consider a fleet of $\numberFleetVeh$ \glspl{abk:av}. In a time-invariant setting, the number of vehicles on arc $\arc \in \setOfArcsRoad$ is expressed as the product of the total vehicles flow on the arc and its travel time. Therefore, we constrain the number of used \glspl{abk:av} as
\begin{equation}
    \label{eq:fleetsizecar}
    \numberFleetUsedVeh = \sum_{\arc \in \setOfArcsRoad} \flowTot{i}{j} \cdot \arcTime 
    \leq \numberFleetVeh.
\end{equation}
\subsection{Discussion}
A few comments are in order.
First, we assume the demand to be time-invariant and allow flows to have fractional values. This assumption is in line with the mesoscopic and system-level planning perspective of our study.
Second, we model congestion effects using a threshold model. This approach can be interpreted as a municipality preventing \glspl{abk:av} to exceed the critical flow density on road arcs. \glspl{abk:av} can be therefore assumed to travel at free-flow speed~\cite{Daganzo2008}. This assumption is realistic for an initial low penetration of \gls{abk:amod} systems in the transportation market, especially when the \gls{abk:av} fleet is of limited size.
Finally, we allow \glspl{abk:av} to transport one customer at the time~\cite{PIM2012}.
\subsection{Co-Design Framework}\label{sec:codesignframework}
We integrate the \gls{abk:iamod} framework presented in \cref{subsec:iamod} in the co-design formalism, allowing one to decompose the \gls{abk:cdpi} of a complex system in the \glspl{abk:dpi} of its individual components in a modular, compositional, and systematic fashion. 
We aim at computing the antichain of resources, quantified in terms of costs, average travel time per trip, and emissions required to provide the mobility service to a set of customers.
In order to achieve this, we decompose the \gls{abk:cdpi} in the \glspl{abk:dpi} of the individual \glspl{abk:av} (\cref{sec:vehdp}), of the \gls{abk:av} fleet (\cref{sec:iamodp}), and of the public transportation system (\cref{sec:subdp}).
The interconnection of the presented \glspl{abk:dpi} is presented in \cref{sec:mcdp}.
\subsubsection{The Autonomous Vehicle Design Problem}
\label{sec:vehdp}
The \gls{abk:av} \gls{abk:dpi} consists of selecting the maximal speed of the \glspl{abk:av}. 
Under the rationale that driving safely at higher speed requires more advanced sensing and algorithmic capabilities, we model the achievable speed of the \glspl{abk:av} $\achievableSpeedVeh$ as a monotone function of the vehicle fixed costs $\costFixVeh$ (resulting from the cost of the vehicle $C_\mathrm{V,v}$ and the cost of its automation $C_\mathrm{V,a}$) and of the mileage-dependent operational costs $\costOpVeh$ (accounting for maintenance, cleaning, energy consumption, depreciation, and opportunity costs~\cite{Mas-ColellWhinstonEtAl1995}).
In this setting, the \gls{abk:av} \gls{abk:dpi} provides the functionality $\achievableSpeedVeh$ and requires the resources $\costFixVeh$ and $\costOpVeh$. Consequently, the functionality space is $\setOfFunctionalities{\mathrm{V}}=\rplusbar$, and the resources space is $\setOfResources{\mathrm{V}}=\rplusbar \times \rplusbar$.
\subsubsection{The Subway Design Problem}
\label{sec:subdp}
We design the public transit infrastructure by means of the service frequency introduced in \cref{subsec:traveltime}. 
Specifically, we assume that the service frequency $\freqTrain$ scales linearly with the size of the train fleet $\numberFleetTrain$ as
\begin{equation}
\freqTrain / \freqTrainBaseline=\numberFleetTrain / \numberFleetTrainBaseline.
\end{equation}
We relate a train fleet of size $\numberFleetTrain$ to the fixed costs $\costFixTrain$ (accounting for train and infrastructural costs) and to the operational costs $\costOpTrain$ (accounting for energy consumption, vehicles depreciation, and train operators' wages).
Given the passenger-independent public transit operation in today's cities, we reasonably assume the operational costs $\costOpTrain$ to be mileage independent and to only vary with the size of the fleet.
Formally, the number of acquired trains $n_\mathrm{S,a}=\numberFleetTrain-\numberFleetTrainBaseline$ is a functionality, whereas $\costFixTrain$ and $\costOpTrain$ are resources. The functionality space is $\setOfFunctionalities{\mathrm{S}}=\nbar$ and the resources space is $\setOfResources{\mathrm{S}}=\rplusbar \times \rplusbar$.
\subsubsection{The I-AMoD Framework Design Problem}
\label{sec:iamodp}
The \gls{abk:iamod} \gls{abk:dpi} considers  demand satisfaction as a functionality. Formally,  $\setOfFunctionalities{\mathrm{O}}=2^{\setOfVertices\times\setOfVertices\times\rplusbar}$ with the partial order $\preceq_{\setOfFunctionalities{\mathrm{O}}}$ defined by $\mathcal{D}_1\coloneqq\{(o^1_i,d^1_i,\alpha^1_i)\}_{i=1}^{M_1}\preceq_{\setOfFunctionalities{\mathrm{O}}}\{(o^2_i,d^2_i,\alpha^2_i)\}_{i=1}^{M_2}\eqqcolon\mathcal{D}_2$ iff for all $(o^1,d^1,\alpha^1)\in\mathcal{D}_1$ there is some $(o^2,d^2,\alpha^2)\in\mathcal{D}_2$ with $o^1=o^2$, $d^1=d^2$, and $\alpha^2_i\geq \alpha^1_i$. In other words, $\mathcal{D}_1\preceq_{\setOfFunctionalities{\mathrm{O}}}\mathcal{D}_2$ if every travel request in $\mathcal{D}_1$ is in $\mathcal{D}_2$ too.
To successfully satisfy a given set of travel requests, we require the following resources:  (i) the achievable speed of the \glspl{abk:av} $\achievableSpeedVeh$, (ii) the number of available \glspl{abk:av} per fleet $\numberFleetVeh$, (iii) the number of trains $n_\mathrm{S,a}$ acquired by the public transportation system, and (iv) the average travel time of a trip
\begin{equation}
    \label{eq:traveltime}
    \averageTravelTime \coloneqq\frac{1}{\alpha_\mathrm{tot}}\cdot \sum_{m\in \mathcal{M}, \arc \in \mathcal{A}} \arcTime \cdot \flow{i}{j},
\end{equation}
with 
$\alpha_\mathrm{tot}\coloneqq \sum_{m\in\mathcal{M}}\alpha_m$,	
(v) the total distance driven by the \glspl{abk:av} per unit time
\begin{equation}
    \distanceVeh \coloneqq\sum_{\arc \in \setOfArcsRoad}\arcLength \cdot \flowTot{i}{j},
\end{equation}
(vi) the total \glspl{abk:av} CO\textsubscript{2} emissions per unit time
   \begin{equation}
    \emissionsVeh \coloneqq\gamma \cdot \sum_{\arc \in \setOfArcsRoad} \arcEnergy \cdot \flowTot{i}{j},
\end{equation}
where $\gamma$ relates the energy consumption and the CO\textsubscript{2} emissions. We assume that customers' trips and \gls{abk:amod} rebalancing strategies are chosen to maximize customers' welfare, defined through the average travel time $\averageTravelTime$. Hence, we link the functionality and resources of the \gls{abk:iamod} \gls{abk:dpi} through the following optimization problem:
\begin{equation}
\begin{split}
\label{eq:TIamodopt}
	\min_{\substack{\flow{\cdot}{\cdot}\geq 0, \\ \flowReba{\cdot}{\cdot} \geq 0}}
	\averageTravelTime &=\frac{1}{\alpha_\mathrm{tot}}\sum_{m\in \mathcal{M}, \arc \in \setOfArcs} \arcTime \cdot \flow{i}{j}\\
	&\mathrm{ s.t. \ Eq. } \eqref{eq:flowconstotal},\
	\mathrm{ Eq. }  \eqref{eq:capacity},\
	\mathrm{ Eq. } \eqref{eq:fleetsizecar}.
\end{split}
\end{equation}
Formally, $\setOfFunctionalities{\mathrm{O}}=\rplusbar$, and $\setOfResources{\mathrm{O}}= \rplusbar \times \nbar \times \nbar \times \rplusbar \times \rplusbar \times \rplusbar$.
\begin{remark}
In general, the optimization problem \eqref{eq:TIamodopt} might possess multiple optimal solutions, making the relation between resources and functionality ill-posed. To overcome this subtlety, if two solutions share the same average travel time, we select the one incurring in the lowest mileage. 	
\end{remark}
\subsubsection{The Monotone Co-Design Problem}
\label{sec:mcdp}
\begin{figure}[tb]
    \begin{center}
    \scalebox{0.45}{
    \begin{tikzpicture}[auto,node distance=0cm]    
    \node [block, scale=1.5,minimum width = 4.5cm,anchor=west] (amod)  {I-AMoD};
    \node [block,below left = 2.5 cm and 1.25cm of amod,  scale=1.5] (veh){Vehicle};
    \node [block,below right = 2.5 cm and 1.25cm of amod, scale=1.5] (sub){Subway};
    \node [nodePre, right = 1.6 cm of veh,name=pre];
    \node [nodePre, left = 2.6 cm of sub,name=pre2];
    \node [nodePre, right = 0.44 cm of pre,name=pre3];
    \node [nodeProd, below = 1 cm of pre3, name=prod];
    \node [nodePre, left = 2.75 cm of prod, name=pre4];
    \node [nodePre, right = 0.4 cm of pre3, name=pre6];
    \node [nodeProd, below = 1.9cm of pre6, name=prod2];
    \node [nodePre, left = 4.7 cm of prod2, name=pre5];
    \node [nodePre, right = 6.4 cm of prod, name=pre7];
    \node [nodeProd, left = 2cm of pre7, name=prod3];
    \node [nodePre, right = 7.5 cm of prod, name=pre8];
    \node [nodePre, below = 1.7 cm of prod3, name=pre9];
    \node [nodePre, below = 0.8 cm of prod2, name=pre10];
    \node [nodeSum, right = 1.1 cm of pre10, name=sum];
    \node [nodePre, below right = 0.6 cm and -0.84cm of amod, name=pre11];
    \node [nodeSum, below right = 0.8cm and 0.2cm of pre11, name=sum2];
    \node [nodePre, right = 0.4 cm of sum2, name=pre12];
    \node [nodeProd, right = 0.4cm of pre12, name=prod4];
    \node [nodePre, right = 0.4 cm of prod4, name=pre13];
    \node [nodePre, below = 1.5 cm of prod, name=pre14];
    \node [nodeSum, right = 0.7 cm of pre13, name=sum3];  
    \node [nodeSum, below = 4.0 cm of veh, name=sum4];
    \node [nodePre, below right = 0.5 cm and 0.5 cm of sum4, name=pre15]; 
    \node [nodeSum, right = 2.0 cm of pre15, name=sum5]; 
    
	\node [nodePreEmpty] at (sum.center|-sum5.center) (pre16) {$\preceq$};
    
    \draw[color=red,thick,dashed] (pre.north) -- node[solid,pos=1,below,circle,color=red,draw,fill=red,scale=0.4] {}(pre.north|-amod.south) node[pos=0.87,left]{$\achievableSpeedVeh$};
    
    \draw[color=DPgreen,thick] (pre.west) -- (veh.east) node[pos=1,right,circle,color=DPgreen,draw,fill=DPgreen,scale=0.4] {};
    
    \draw[color=red,thick,dashed] (pre2.north) -- (pre2|-amod.south) node[pos=0.87,right]{$n_\mathrm{S,a}$}  node[solid,pos=1,below,circle,color=red,draw,fill=red,scale=0.4]{};
    
    \draw[color=DPgreen,thick] (pre2.east) -- (sub.west) node[pos=1,left,circle,color=DPgreen,draw,fill=DPgreen,scale=0.4] {};
    
    \draw[color=red,thick,dashed] (pre3.north) -- (pre3|-amod.south) node[pos=0.87,left]{$\distanceVeh$}  node[solid,pos=1,below,circle,color=red,draw,fill=red,scale=0.4] {};
    
    \draw[color=DPgreen,thick] (pre3.south) -- (prod.north) node[pos=1,above,circle,color=DPgreen,draw,fill=DPgreen,scale=0.4,solid] {};
    
    \draw[color=red,thick,dashed] (prod.south) -- (pre14.north) node[pos=0,below,circle,color=red,draw,fill=red,scale=0.4,solid] {};
    
    \draw [-, color=red,thick,dashed] (pre4|-veh.south) -- (pre4.north) node[solid,pos=0,below,circle,color=red,draw,fill=red,scale=0.4]{} node[pos=0.5,right]{$\costOpVeh$};
    
    \draw [-, color=red,thick,dashed] (pre5|-veh.south) -- (pre5.north) node[solid,pos=0,below,circle,color=red,draw,fill=red,scale=0.4]{} node[pos=0.2,right]{$\costFixVeh$} ;
    
    \node[rectangle,align=center] at ($(pre5)+(-1.7,-1.5)$) (pre5Label) {co-design  \\ constraint};
    \draw[->,thick]  (pre5Label) to[out=60,in=135] (pre5);
    
    \draw [-, color=DPgreen,thick]  (pre4.east) --  (prod.west) node[pos=1,left,circle,color=DPgreen,draw,fill=DPgreen,scale=0.4] {};
    
    \draw [-, color=red,thick,dashed] (pre8|-sub.south) -- (pre8.north) node[solid,pos=0,below,circle,color=red,draw,fill=red,scale=0.4]{} node[pos=0.5,right]{$\costOpTrain$};
    
    \draw [-, color=red,thick,dashed] (pre7|-sub.south) -- (pre7.north) node[solid,pos=0,below,circle,color=red,draw,fill=red,scale=0.4]{} node[pos=0.5,right]{$\costFixTrain$};
    
    \draw[color=red,thick,dashed] (pre6.north) -- (pre6|-amod.south) node[pos=0.87,right]{$n_\mathrm{V,max}$}  node[solid,pos=1,below,circle,color=red,draw,fill=red,scale=0.4] {};
    
    \draw[color=DPgreen,thick] (pre5.east) -- (prod2.west) node[pos=1,left,circle,color=DPgreen,draw,fill=DPgreen,scale=0.4] {};
    
    \draw[color=DPgreen,thick] (pre6.south) -- (prod2.north) node[pos=1,above,circle,color=DPgreen,draw,fill=DPgreen,scale=0.4] {};
    
    \draw[color=DPgreen,thick] (pre2.east) -| (prod3.north) node[pos=1,above,circle,color=DPgreen,draw,fill=DPgreen,scale=0.4] {};
    
    \draw[color=DPgreen,thick] (prod3.east) -- (pre7.west) {} node[solid,pos=0,right,circle,color=DPgreen,draw,fill=DPgreen,scale=0.4]{};
    
    \draw[color=red,thick,dashed] (prod2.south) -- (pre10.north) node[pos=0,below,circle,color=red,draw,fill=red,scale=0.4,solid] {};
    
    \draw[color=red,thick,dashed] (prod3.south) -- (pre9.north) node[pos=0,below,circle,color=red,draw,fill=red,scale=0.4,solid] {};
    
    \draw[color=DPgreen,thick] (pre10.east) -- (sum.west) node[pos=1,left,circle,color=DPgreen,draw,fill=DPgreen,scale=0.4] {};
    
    \draw[color=DPgreen,thick] (pre9.west) -- (sum.east) node[pos=1,right,circle,color=DPgreen,draw,fill=DPgreen,scale=0.4] {};
           
    \draw[color=DPgreen,thick] (pre8.south) |- (sum4.east)  node[pos=1,right,circle,color=DPgreen,draw,fill=DPgreen,scale=0.4] {};
    
    \draw[color=DPgreen,thick] (pre14.west) -| (sum4.north)  node[pos=1,circle,above,color=DPgreen,draw,fill=DPgreen,scale=0.4] {};
    
    \draw[color=red,thick,dashed] (sum4.south) |- (pre15.west) node[solid,pos=0,below,circle,color=red,draw,fill=red,scale=0.4]{};
    
    \draw[color=DPgreen,thick] (pre15.east) -- (sum5.west) node[pos=1,left,circle,color=DPgreen,draw,fill=DPgreen,scale=0.4] {};
    
    \draw[color=red,thick,dashed]  (sum5.south) --++ (0,-0.8) node[solid,pos=0,below,circle,color=red,draw,fill=red,scale=0.4]{} node[pos=0.8,right]{$C_\mathrm{tot}$};
    
    \draw[color=red,thick,dashed] (sum.south) -- (pre16.north) {}node[pos=0,below,circle,color=red,draw,fill=red,scale=0.4,solid] {};

    \draw[color=DPgreen,thick] (sum5.east) -- (pre16.west) node[pos=0,right,circle,color=DPgreen,draw,fill=DPgreen,scale=0.4] {};
    
    \draw[color=DPgreen,thick] (amod.north) --++ (0,1.25) node[pos=0.8,right]{$\alpha_\mathrm{tot}$}  node[pos=0,above,circle,color=DPgreen,draw,fill=DPgreen,scale=0.4] {};
    
    \draw[color=DPgreen,thick] (pre2.east) -| (sum2.south) {}node[solid,pos=1,below,circle,color=DPgreen,draw,fill=DPgreen,scale=0.4]{};
    
    \draw[color=red,thick,dashed] (pre11.north) -- (pre11|-amod.south) node[pos=0.5,right]{$\emissionsVeh$}  node[solid,pos=1,below,circle,color=red,draw,fill=red,scale=0.4] {};
    
    \draw[color=DPgreen,thick] (sum2.west) --++ (-0.75,0) node[pos=0,left,circle,color=DPgreen,draw,fill=DPgreen,scale=0.4] {} node[pos=0.7,below]{$\numberFleetTrain$};
    
    \draw[color=DPgreen,thick] (prod4.north) --++ (0,0.5) node[pos=0,above,circle,color=DPgreen,draw,fill=DPgreen,scale=0.4] {}node[pos=0.7,right]{$\emissionsSub$};
    
    \draw[color=red,thick,dashed] (sum2.east) -- (pre12.west) node[pos=0,right,circle,color=red,draw,fill=red,scale=0.4,solid] {};
    
    \draw[color=DPgreen,thick] (pre12.east) -- (prod4.west) node[pos=1,left,circle,color=DPgreen,draw,fill=DPgreen,scale=0.4] {};
    
    \draw[color=red,thick,dashed] (prod4.east) -- (pre13.west) node[pos=0,right,circle,color=red,draw,fill=red,scale=0.4,solid] {};
    
    \draw[color=DPgreen,thick] (pre13.east) -- (sum3.west) node[pos=1,left,circle,color=DPgreen,draw,fill=DPgreen,scale=0.4] {};
    
    \draw[color=DPgreen,thick] (pre11.east) -| (sum3.north) node[pos=1,above,circle,color=DPgreen,draw,fill=DPgreen,scale=0.4] {};
    
    \draw[color=red,thick,dashed] (sum3.south) --++ (0,-5.9) node[pos=0,below,circle,color=red,draw,fill=red,scale=0.4,solid] {} -|++ (-3,-1.8) node[pos=0.95,right]{$\emissionsTot$};
    
    \draw[color=red, thick,dashed] ($(amod.south)+(0.75,0)$) --++ (0,-5.5) node[solid,pos=0,below,circle,color=red,draw,fill=red,scale=0.4] {}  -|++ (1,-4.45) node[pos=0.97,right]{$\averageTravelTime$};
    
    \draw[color=DPgreen,thick] (prod2.east) --++ (0.75,0) node[pos=0,right,circle,color=DPgreen,draw,fill=DPgreen,scale=0.4] {} node[pos=0.5,above,DPgreen]{$\lifeVeh$};
    
    \draw[color=DPgreen,thick] (prod3.west) --++ (-0.75,0) node[pos=0,left,circle,color=DPgreen,draw,fill=DPgreen,scale=0.4] {} node[pos=0.5,above,DPgreen]{$\lifeTrain$};
    
    \node[rectangle,align=center,color=red] at ($(1.7,-11)$) (name) {total cost};
    \node[rectangle,align=center,color=red] at ($(5.1,-11.3)$) (name2) {average \\ travel time};
    \node[rectangle,align=center,color=red] at ($(8,-11.25)$) (name3) {total \\ emissions};
    \node[rectangle,align=center,color=DPgreen] at ($(3.4,2.5)$) (name3) {total \\ request rate};
    
    \draw[dashed] (-3.6,-9.9) rectangle (11.7,1.2);
    \end{tikzpicture}}
    \caption{Schematic representation of the \gls{abk:cdpi}. In solid \textcolor{DPgreen}{green} the provided functionalities and in dashed \textcolor{red}{red} the required resources. The edges represent co-design constraints: The resources required by a first design problem are the lower bound for the functionalities provided by the second one.}
    \label{fig:mcdp}  
    \vspace{-6mm}
    \end{center}
\end{figure}
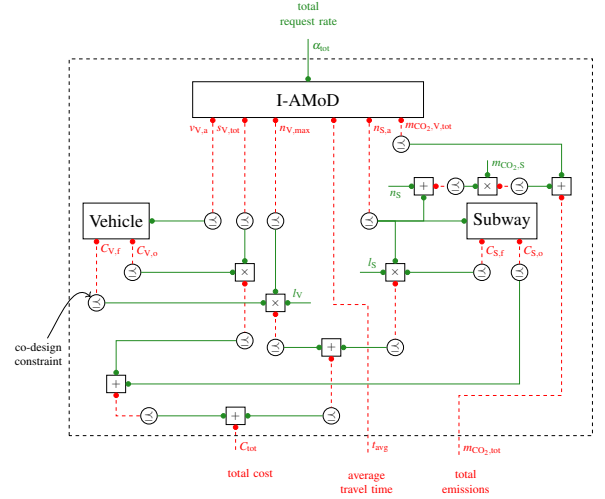

The functionality of the system is to provide mobility service to the customers. Formally, the functionality provided by the \gls{abk:cdpi} is the set of travel requests. To provide the mobility service, the following three resources are required.
First, on the customers' side, we require an average travel time, defined in~\eqref{eq:traveltime}.
Second, on the municipality side, the resource is the total transportation cost of the intermodal mobility system. Assuming an average vehicles' life $\lifeVeh$, an average trains' life  $\lifeTrain$, and a baseline subway fleet of $\numberFleetTrainBaseline$ trains, we express the total costs as
\begin{equation}
    \costTot=\costVeh + \costSub,
\end{equation}
where $\costVeh$ is the \gls{abk:av}-related cost
\begin{equation}
    \costVeh=\frac{\costFixVeh}{\lifeVeh}\cdot \numberFleetVeh + \costOpVeh \cdot \distanceVeh,
\end{equation}
and $\costSub$ is the public transit-related cost
\begin{equation}
    \costSub=\frac{\costFixTrain}{\lifeTrain}\cdot n_\mathrm{S,a} + \costOpTrain.
\end{equation}
Third, on the environmental side, the resources are the total CO\textsubscript{2} emissions
\begin{equation}
    \emissionsTot =\emissionsVeh + \emissionsSub \cdot \numberFleetTrain,
\end{equation}
where $\emissionsSub$ represents the CO\textsubscript{2} emissions of a single train.
Formally, the set of travel requests $\{\rho_m\}_{m\in\cM}$ is the \gls{abk:cdpi} functionality, whereas $\averageTravelTime$, $\costTot$, and $\emissionsTot$ are its resources. Consistently, the functionality space is $\setOfFunctionalities{}=\rplusbar$ and the resources space is $\setOfResources{}=\rplusbar \times \rplusbar \times \rplusbar$. Note that the resulting \gls{abk:cdpi} (\cref{fig:mcdp}) is indeed monotone, since it consists of the interconnection of monotone \glspl{abk:dpi}~\cite{Censi2015}.
\subsubsection{Discussion}
A few comments are in order.
First, we lump the autonomy functionalities in its achievable velocity. We leave to future research more elaborated \gls{abk:av} models, accounting for instance for accidents rates~\cite{Richards2010} and for safety levels. 
Second, we assume the service frequency of the subway system to scale linearly with the number of trains. We inherently rely on the assumption that the existing infrastructure can homogeneously accommodate the acquired train cars. To justify the assumption, we include an upper bound on the number of potentially acquirable trains in our case study design in \cref{sec:results}.
Third, we highlight that the \gls{abk:iamod} framework is only one of the many feasible ways to map total demand to travel time, costs, and emissions. Specifically, practitioners can easily replace the corresponding \gls{abk:dpi} with more sophisticated models (e.g., simulation-based frameworks like AMoDeus~\cite{Ruch2018}), as long as the monotonicity of the \gls{abk:dpi} is preserved. In our setting, we conjecture the customers' and vehicles' routes to be centrally controlled by the municipality in a socially-optimal fashion. Implicitly, we rely on the existence of effective incentives aligning private and societal interests. The study of such incentives represents an avenue for future research.
Fourth, we assume a homogenous fleet of \glspl{abk:av}. Nevertheless, our model is readily extendable to capture heterogeneous fleets. Finally, we consider a fixed travel demand, and compute the antichain of resources providing it. Nonetheless, our formalization can be readily extended to arbitrary demand models preserving the monotonicity of the \gls{abk:cdpi} (accounting for instance for elastic effects). We leave this topic to future research.

\section{Results}
\label{sec:results}
In this section, we leverage the framework presented in \cref{sec:codesignav} to perform a real-world case study of Washington D.C., USA. \cref{sec:case1} details the case study. We then present numerical results in \cref{sec:case2,sec:case3}.

\subsection{Case Study}
\label{sec:case1}
We base our studies on a real-world case of the urban area of Washington D.C., USA.
We import the road network and its features from OpenStreetMap~\cite{HaklayWeber2008}. 
The public transit network and its schedules are extracted from the GTFS data~\cite{GTFS2019}.
Demand data is obtained by merging the origin-destination pairs of the morning peak of May 31st 2017 provided by taxi companies~\cite{ODDC2017} and the Washington Metropolitan Area Transit Authority (WMATA)~\cite{PIM2012}.
Given the lack of reliable demand data for the MetroBus system, we focus our studies on the MetroRail system and its design, inherently assuming  MetroBus commuters to be unaffected by our design methodology. To conform with the large presence of ride-hailing companies, we scale the taxi demand rate by factor of 5~\cite{Siddiqui2018b}. Overall, the demand dataset includes 15,872 travel requests, corresponding to a demand rate of \unitfrac[24.22]{requests}{s}.
To account for congestion effects, we compute the nominal road capacity as in~\cite{DoA1977} and assume an average baseline road usage of 93\%, in line with~\cite{DixonIrshadEtAl2018}. We summarize the main parameters together with their bibliographic sources in \cref{tab:params}. In the remainder of this section, we tailor and solve the co-design problem presented in \cref{sec:codesignav} through the PyMCDP solver~\cite{Censi2019}, and investigate the influence of different \glspl{abk:av} costs on the design objectives and strategies.

\begin{table}[tb]
	\begin{center}
		\begin{tiny}
		\addtolength{\tabcolsep}{-0.04cm}
		\begin{tabular}{llllccccc}
			\toprule
			& \multicolumn{2}{l}{\textbf{Parameter}} & \textbf{Name}  & \multicolumn{3}{c}{\textbf{Value}} &\textbf{Units}& \textbf{Source}\\
			\midrule
			& Road usage & & $u_{ij}$ &\multicolumn{3}{c}{93} &\unit[]{\%}& ~\cite{DixonIrshadEtAl2018}\\
			\midrule
			\multirow{15}{*}{\rotatebox[origin=c]{90}{Vehicle}} &
			&&& \textbf{C1} & \textbf{C2.1} & \textbf{C2.2}\\ \cline{4-8}  \\[-1.0em]
			&\multicolumn{2}{l}{Operational cost} & $\costOpVeh$ &0.084 & 0.084&0.062 &\unitfrac[]{USD}{mile}&~\cite{PavlenkoSlowikEtAl2019, BoeschBeckerEtAl2018}\\
			&\multicolumn{2}{l}{Cost} & $\costVeh$ & 32,000&32,000&26,000&\unitfrac[]{ USD}{car}& ~\cite{PavlenkoSlowikEtAl2019}\\
			&\multirow{7}{*}{Automation cost} & \unit[20]{mph} & \multirow{7}{*}{$C_\mathrm{V,a}$} &15,000&20,000&3,700&\unitfrac[]{USD}{car}& ~\cite{BoeschBeckerEtAl2018,FagnantKockelman2015,BauerGreenblattEtAl2018,Wadud2017,Litman2019}\\
			 && \unit[25]{mph} &  &15,000&30,000&4,400& \unitfrac[]{USD}{car}&~\cite{BoeschBeckerEtAl2018,FagnantKockelman2015,BauerGreenblattEtAl2018,Wadud2017,Litman2019}\\
			 && \unit[30]{mph} &  &15,000&55,000&6,200&\unitfrac[]{USD}{car}& ~\cite{BoeschBeckerEtAl2018,FagnantKockelman2015,BauerGreenblattEtAl2018,Wadud2017,Litman2019}\\
			 && \unit[35]{mph} &  &15,000&90,000&8,700&\unitfrac[]{USD}{car}& ~\cite{BoeschBeckerEtAl2018,FagnantKockelman2015,BauerGreenblattEtAl2018,Wadud2017,Litman2019}\\
			 && \unit[40]{mph} &  &15,000&115,000&9,800&\unitfrac[]{USD}{car}&~\cite{BoeschBeckerEtAl2018,FagnantKockelman2015,BauerGreenblattEtAl2018,Wadud2017,Litman2019}\\
			 && \unit[45]{mph} &  &15,000&130,000&12,000& \unitfrac[]{USD}{car}&~\cite{BoeschBeckerEtAl2018,FagnantKockelman2015,BauerGreenblattEtAl2018,Wadud2017,Litman2019}\\
			 && \unit[50]{mph} &  &15,000&150,000&13,000&\unitfrac[]{USD}{car}& ~\cite{BoeschBeckerEtAl2018,FagnantKockelman2015,BauerGreenblattEtAl2018,Wadud2017,Litman2019}\\
			&\multicolumn{2}{l}{Vehicle life} &$\lifeVeh$ &5 & 5& 5 &\unit[]{years}&~\cite{PavlenkoSlowikEtAl2019}\\
			&\multicolumn{2}{l}{CO$_2$ per Joule}& $\gamma$ & 0.14&0.14&0.14&\unitfrac[]{g}{kJ}& ~\cite{Watttime2018}\\
			&\multicolumn{2}{l}{Time $\mathcal{G}_\mathrm{W}$ to $\mathcal{G}_\mathrm{R}$ } & $\timePedestrianRoad$ & 300&300&300&\unit[]{s}&-\\
			&\multicolumn{2}{l}{Time $\mathcal{G}_\mathrm{R}$ to $\mathcal{G}_\mathrm{W}$} & $\timeRoadPedestrian$ & 60&60&60&\unit[]{s}&-\\
			&\multicolumn{2}{l}{Speed fraction}&$\beta$&$\frac{1}{1.3}$&$\frac{1}{1.3}$&$\frac{1}{1.3}$&\unit[]{-}&-\\
			\midrule
			\multirow{10}{*}{\rotatebox[origin=c]{90}{Public transit}}
			&\multirow{3}{*}{Operational cost} & \unit[100]{\%}& \multirow{3}{*}{$\costOpTrain$} & \multicolumn{3}{c}{148,000,000} &\unitfrac[]{USD}{year}&~\cite{WMATA2017}\\
			&& \unit[133]{\%}&  & \multicolumn{3}{c}{197,000,000} &\unitfrac[]{USD}{year}& ~\cite{WMATA2017}\\
		    && \unit[200]{\%}&  & \multicolumn{3}{c}{295,000,000} &\unitfrac[]{USD}{year}& ~\cite{WMATA2017}\\			
			&\multicolumn{2}{l}{Fixed cost} & $\costFixTrain$ &  \multicolumn{3}{c}{14,500,000}  &\unitfrac[]{USD}{train}& ~\cite{Aratani2015}\\
			&\multicolumn{2}{l}{Train life}& $\lifeTrain$ & \multicolumn{3}{c}{30} &{\unit[]{years}}& ~\cite{Aratani2015}\\ 
			&\multicolumn{2}{l}{Emissions/train}& $\emissionsSub$ &\multicolumn{3}{c}{140,000} &\unitfrac[]{kg}{year}&~\cite{WMATA2018} \\
			&\multicolumn{2}{l}{Fleet baseline} & $n_\mathrm{S,base}$ &\multicolumn{3}{c}{112}&\unit[]{trains}& ~\cite{Aratani2015}\\
			&\multicolumn{2}{l}{Service frequency}& $\varphi_{j,\mathrm{base}}$ & \multicolumn{3}{c}{$1/6$}&\unitfrac[]{1}{min}& ~\cite{Jaffe2015}\\
			&\multicolumn{2}{l}{Time $\mathcal{G}_\mathrm{W}$ to $\mathcal{G}_\mathrm{P}$} & $\timePedestrianSubway$& \multicolumn{3}{c}{$60$}&\unit[]{s}&-\\
			&\multicolumn{2}{l}{Time $\mathcal{G}_\mathrm{P}$ to $\mathcal{G}_\mathrm{W}$} & $\timeSubwayPedestrian$ & \multicolumn{3}{c}{$60$}&\unit[]{s}&-\\
		\bottomrule
	\end{tabular}
	\caption{Parameters, variables, numbers, and units for the case studies.}
	\label{tab:params}
	\end{tiny}
	\vspace{-6mm}
	\end{center}
\end{table}

\subsection{Case 1 - Constant Cost of Automation}
\label{sec:case2}
\begin{figure}[tbh]
    \begin{center}
    \begin{subfigure}[h]{\columnwidth}
    \includegraphics[width=\columnwidth]{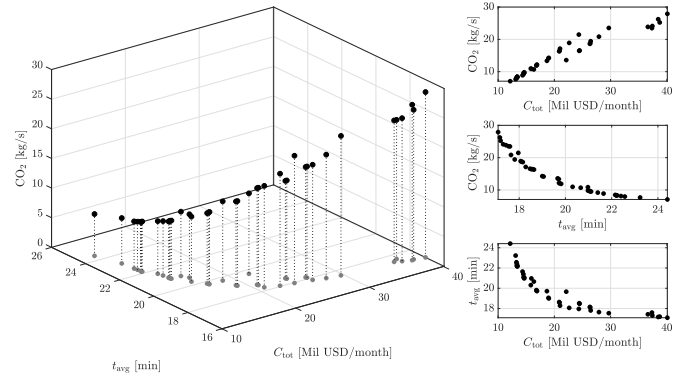}
    \caption{Left: Three-dimensional representation of antichain elements and their projection in the cost-time space. Right: Two-dimensional projections.}
    \label{fig:resultsart3D}
    \end{subfigure}
    ~
    \begin{subfigure}[h]{\columnwidth}
	\includegraphics[width=\columnwidth]{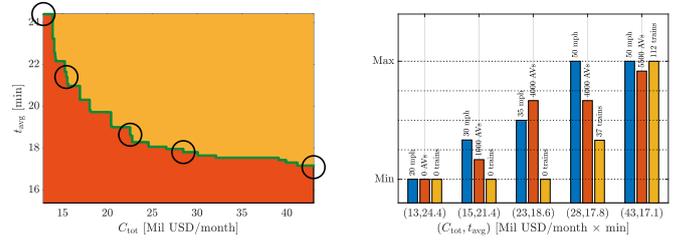}
    \caption{Results for constant automation costs. On the left, the two-dimensional representation of the antichain elements: In red are the unfeasible strategies, in orange the feasible but irrational solutions, and in green the Pareto front. On the right, the implementations corresponding to the highlighted antichain elements, quantified in terms of achievable vehicle speed, \gls{abk:av} fleet size, and train fleet size.}
    \label{fig:resultsart2D}
    \end{subfigure}
    \caption{Solution of the \gls{abk:cdpi}: state-of-the art case.}
    \label{fig:resultsart}
    \vspace{-6mm}
    \end{center}
\end{figure}

In line with~\cite{BoeschBeckerEtAl2018,FagnantKockelman2015,BauerGreenblattEtAl2018,Wadud2017, Litman2019}, we first assume an average achievable-velocity-independent cost of automation. As discussed in \cref{sec:codesignav}, we design the system by means of subway service frequency, \gls{abk:av} fleet size, and achievable free-flow speed. Specifically, we allow the municipality to (i) increase the subway service frequency $\freqTrain$ by a factor of 0\%, 33\%, or 100\%, (ii) deploy an \glspl{abk:av} fleet of size $\numberFleetVeh \in \{0,500,1000,\hdots,6000\}$ vehicles, and (iii) design the single \gls{abk:av} achievable velocity $\achievableSpeedVeh\in \{\unit[20]{mph},\unit[25]{mph},\hdots,\unit[50]{mph}\}$. 
We assume the \glspl{abk:av} fleet to be composed of battery electric BEV-250 mile \glspl{abk:av}~\cite{PavlenkoSlowikEtAl2019}.
In \cref{fig:resultsart3D}, we show the solution of the co-design problem by reporting the antichain consisting of the total transportation cost, average travel time, and total CO\textsubscript{2} emissions. These solutions are \emph{rational} (and not comparable) in the sense that there exists no instance which simultaneously yields lower cost, average travel time, and emissions. 
For the sake of clarity, we opt for a two-dimensional antichain representation, by translating and including the emissions in the total cost. To do so, we consider the conversion factor \unitfrac[40]{USD}{kg}~\cite{HowardSylvan2015}. Note that since this transformation preserves the monotonicity of the \gls{abk:cdpi} it smoothly integrates in our framework.
Doing so, we can conveniently depict the co-design strategies through the two-dimensional antichain (\cref{fig:resultsart2D}, left) and the corresponding municipality actions (\cref{fig:resultsart2D}, right).
Generally, as the municipality budget increases, the average travel time per trip required to satisfy the given demand decreases, reaching a minimum of about \unit[17.1]{min} with an expense of around \unitfrac[43]{Mil USD}{month}.
This configuration corresponds to a fleet of \unit[5,500]{\glspl{abk:av}} able to drive at \unit[50]{mph} and to the doubling of the current MetroRail train fleet.
On the other hand, the smallest rational investment of \unitfrac[12.9]{Mil USD}{month} leads to a \unit[42]{\%} higher average travel time, corresponding to a inexistent autonomous fleet and an unchanged subway infrastructure. Notably, an expense of \unitfrac[23]{Mil USD}{month} (\unit[48]{\%} lower than the highest rational investment) only increases the minimal required travel time by \unit[9]{\%}, requiring a fleet of \unit[4,000]{vehicles} able to drive at \unit[35]{mph} and no acquisition of trains. Conversely, an investment of \unitfrac[15.6]{Mil USD}{month} (just \unitfrac[2]{Mil USD}{month} more than the minimal rational investment) provides a \unit[3]{min} shorter travel time. 
Remarkably, the design of \glspl{abk:av} able to exceed \unit[40]{mph} only improves the average travel time by \unit[6]{\%}, and it is rational just starting from an expense of \unitfrac[22.8]{Mil USD}{month}.
This suggests that the design of faster vehicles mainly results in higher emission rates and costs, without substantially contributing to a more time-efficient demand satisfaction. 
Finally, it is rational to improve the subway system only starting from a budget of \unitfrac[28.5]{Mil USD}{month}, leading to a travel time improvement of just \unit[4]{\%}. This trend can be explained with the high train acquisition and increased operation costs, related to the subway reinforcement. We expect this phenomenon to be more marked for other cities, considering the moderate operation costs of the MetroRail subway system due to its automation~\cite{Jaffe2015} and related benefits~\cite{WangZhangEtAl2016}.

\subsection{Case 2 - Speed-Dependent Automation Costs}
\label{sec:case3}
To relax the potentially unrealistic assumption of a velocity-independent automation cost, we consider a performance-dependent cost structure. The large variance in sensing technologies and their reported performances~\cite{GawronKeoleianEtAl2018} suggests that this rationale is reasonable. Indeed, the technology required today to safely operate an autonomous vehicle at \unit[50]{mph} is substantially more sophisticated, and therefore more expensive, than the one needed at \unit[20]{mph}. To this end, we adopt the cost structure reported in \cref{tab:params}. Furthermore, the frenetic evolution of automation techniques intricates their monetary quantification. Therefore, we perform our studies with current (2020) costs as well as with their projections for the upcoming decade (2025)~\cite{Lienert2019,PavlenkoSlowikEtAl2019}.

\subsubsection{Case 2.1 - 2020}
\begin{figure}[tb]
    \begin{center}
    \begin{subfigure}[h]{\columnwidth}
    \includegraphics[width=\columnwidth]{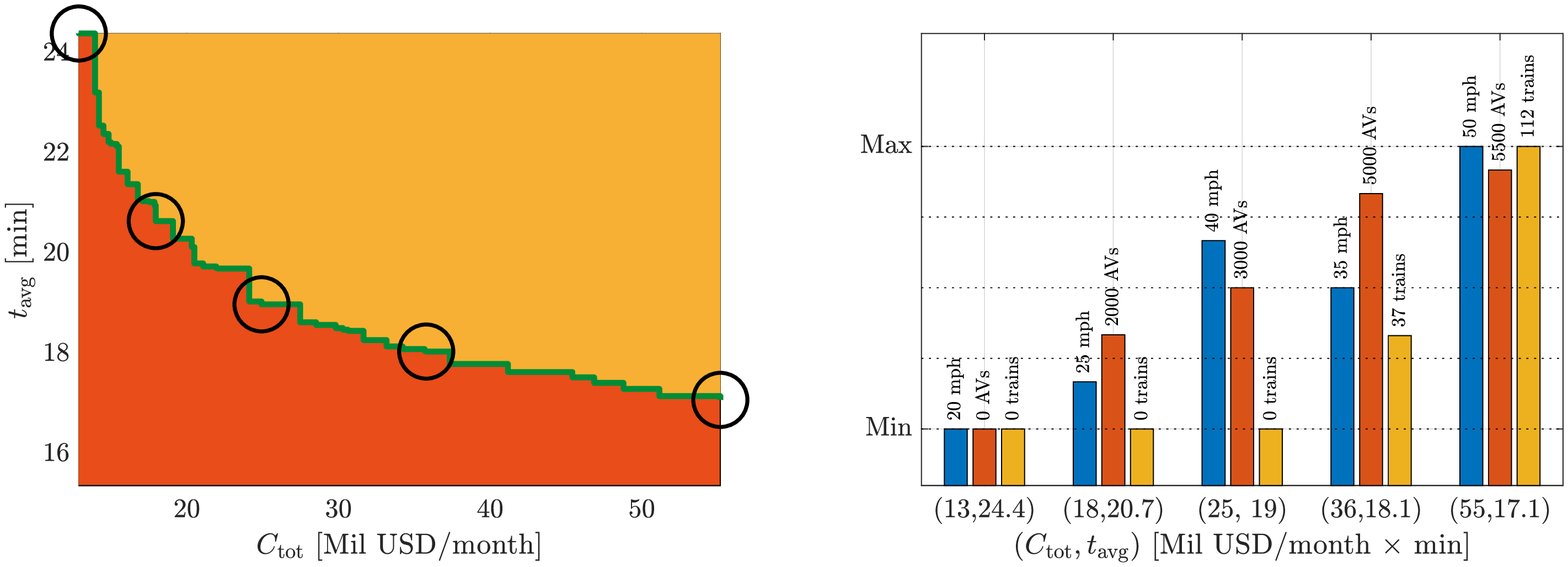}
    \caption{Results for speed-dependent automation costs in 2020.}
    \label{fig:results19}
     \end{subfigure}
     ~
     \begin{subfigure}[h]{\columnwidth}
    \includegraphics[width=\columnwidth]{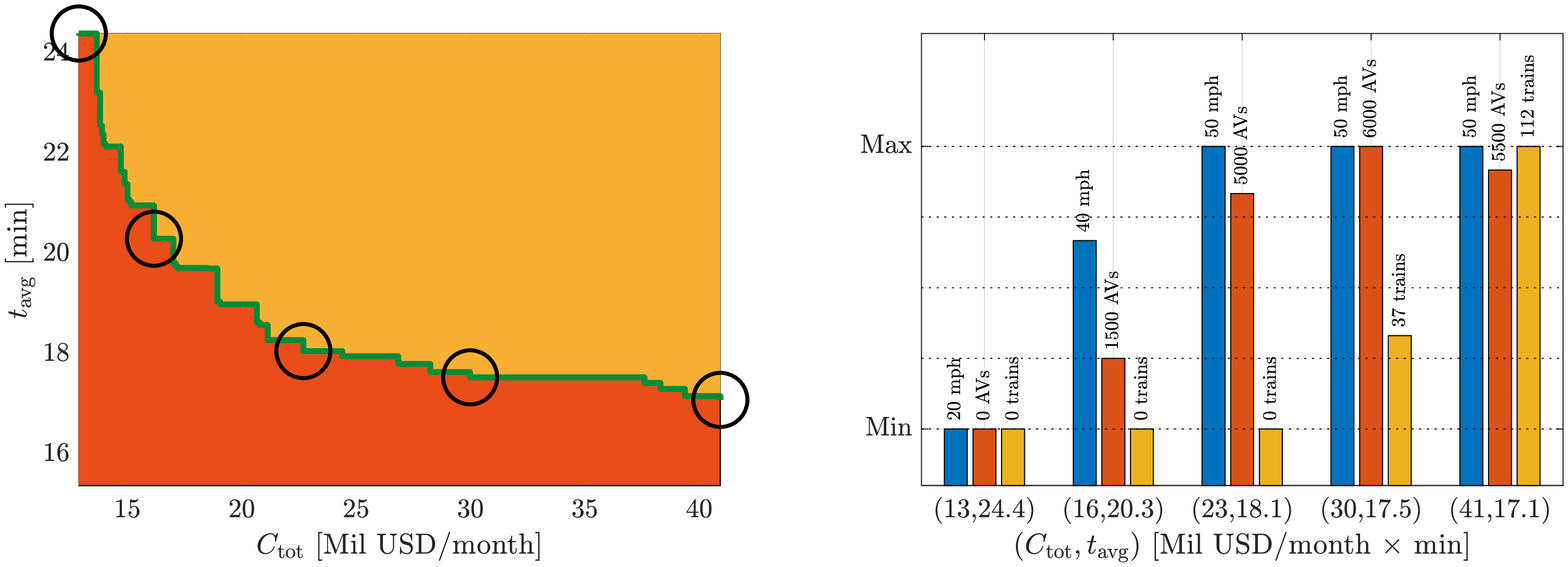}
    \caption{Results for speed-dependent automation costs in 2025.}
    \label{fig:results25}
     \end{subfigure}
    \caption{Results for the speed-dependent automation costs. On the left, the two-dimensional representation of the antichain elements: In red are the unfeasible strategies, in orange the feasible but irrational solutions, and in green the Pareto front. On the right, the implementations corresponding to the highlighted antichain elements.}
    \label{fig:results1925}
    \vspace{-6mm}
    \end{center}
\end{figure}

We study the hypothetical case of an immediate \gls{abk:av} fleet deployment. We introduce the aforementioned velocity-dependent automation cost structure and obtain the results reported in \cref{fig:results19}. Comparing these results with the state-of-the-art parameters presented in \cref{fig:resultsart} confirms the previously observed trend concerning high vehicle speeds. Indeed, spending \unitfrac[24.9]{Mil USD}{month} (\unit[55]{\%} lower than the highest rational expense) only increases the average travel time by \unit[10]{\%}, requiring a fleet of \unit[3,000]{\glspl{abk:av}} at \unit[40]{mph} and no subway interventions. Nevertheless, the comparison shows two substantial differences.
First, the budget required to reach the minimum travel time of \unit[17.1]{min} is \unit[28]{\%} higher compared to the previous case, and consists of the same strategy for the municipality, i.e., doubling the train fleet and having a fleet of \unit[5,500]{\glspl{abk:av}} at \unit[50]{mph}.
Second, the higher vehicle costs result in an average \gls{abk:av} fleet growth of \unit[5]{\%}, an average velocity reduction of \unit[9]{\%}, and an average train fleet growth of \unit[7]{\%}. This trend suggests that, compared to Case 1, rational design strategies foster larger fleets and less performing \glspl{abk:av}.
\subsubsection{Case 2.2 - 2025}
Experts forecast a substantial decrease of automation costs (up to \unit[90]{\%}) in the next decade, mainly due to mass-production of the \glspl{abk:av} sensing technology~\cite{Lienert2019, WCP2018}.
In line with this prediction, we inspect the futuristic scenario by solving the \gls{abk:cdpi} for the adapted automation costs, and report the results in \cref{fig:results25}.
Two comments are in order. First, the maximal rational budget is \unit[25]{\%} lower than in the immediate adoption case. Second, the reduction in autonomy costs clearly eases the acquisition of more performant \glspl{abk:av}, increasing the average vehicle speed by \unit[10]{\%}. As a direct consequence, the \gls{abk:av} and train fleets are reduced in size by \unit[5]{\%} and \unit[10]{\%}, respectively. 

\subsection{Discussion}
We conclude the analysis of our case study with two final comments.
First, the presented case studies illustrate the ability of our framework to extract the set of rational design strategies for an \gls{abk:av}-enabled mobility system.
This way, stakeholders such as \glspl{abk:av} companies, transportation authorities, and policy makers can get transparent and interpretable insights on the impact of future interventions. 
Second, we perform a sensitivity analysis through the variation of the autonomy cost structures.
On the one hand, this reveals a clear transition from small fleets of fast \glspl{abk:av} (in the case of low autonomy costs) to slow fleets of numerous \glspl{abk:av} (in the case of high autonomy costs).
On the other hand, our studies highlight that investments in the public transit infrastructure are rational only when large budgets are available.
Indeed, the onerous train acquisition and operation costs lead to a comparative advantage of \gls{abk:av}-based mobility.

\section{Conclusion}\label{sec:conclusion}
In this paper, we leveraged the mathematical theory of co-design to propose a design framework for \gls{abk:av}-enabled mobility systems. Specifically, the nature of our framework allows both for the modular and compositional interconnection of the \glspl{abk:dpi} of different mobility options and for multiple objectives.
Starting from the multi-commodity flow model of an \gls{abk:iamod} system, we optimize the design of \glspl{abk:av} and public transit both from a vehicle-centric and fleet-level perspective. In particular, we studied the problem of deploying a fleet of \glspl{abk:av} providing on-demand mobility in cooperation with public transit, optimizing the speed achievable by the vehicles, the fleet size, and the service frequency of the subway lines. 
Our framework allows the stakeholders involved in the mobility ecosystem, from vehicle developers all the way to mobility-as-a-service companies and governmental authorities, to characterize rational trajectories for technology and investment development. We showcased our methodology on a real-world case study of Washington D.C., USA. Notably, our problem formulation allows for a systematic analysis of incomparable objectives, providing stakeholders with analytical insights for the socio-technical design of \gls{abk:av}-enabled mobility systems.
This work opens the field for the following future research streams:\\
\emph{Modeling:} First, we would like to extend the presented framework to capture additional modes of transportation, such as micromobility, and heterogeneous fleets with different self-driving infrastructures, propulsion systems, and passenger capacity. Second, we would like to investigate variable demand models. Finally, we would like to analyze the interactions between multiple stakeholders, characterizing the equilibrium arising from their conflicting interests.\\
\emph{Algorithms:} It is of interest to tailor co-design algorithmic frameworks to the particular case of transportation \glspl{abk:dpi}, possibly leveraging their specific structure.\\
\emph{Application:} Finally, we would like to devise a user-friendly web interface which supports mobility stakeholders to reason about strategic interventions in urban areas.

\bibliographystyle{IEEEtran}
\bibliography{paper}

\begin{thebibliography}{10}
\providecommand{\url}[1]{#1}
\csname url@rmstyle\endcsname
\providecommand{\newblock}{\relax}
\providecommand{\bibinfo}[2]{#2}
\providecommand\BIBentrySTDinterwordspacing{\spaceskip=0pt\relax}
\providecommand\BIBentryALTinterwordstretchfactor{4}
\providecommand\BIBentryALTinterwordspacing{\spaceskip=\fontdimen2\font plus
\BIBentryALTinterwordstretchfactor\fontdimen3\font minus
  \fontdimen4\font\relax}
\providecommand\BIBforeignlanguage[2]{{%
\expandafter\ifx\csname l@#1\endcsname\relax
\typeout{** WARNING: IEEEtran.bst: No hyphenation pattern has been}%
\typeout{** loaded for the language `#1'. Using the pattern for}%
\typeout{** the default language instead.}%
\else
\language=\csname l@#1\endcsname
\fi
#2}}

\bibitem{ZardiniLanzettiEtAl2020}
G.~Zardini, N.~Lanzetti, M.~Salazar, A.~Censi, E.~Frazzoli, and M.~Pavone,
  ``Towards a co-design framework for future mobility systems,'' in
  \emph{{Annual Meeting of the Transportation Research Board}}, Washington
  D.C., United States, 2020.

\bibitem{Censi2015}
A.~Censi, ``A mathematical theory of co-design,'' \emph{arXiv preprint
  arXiv:1512.08055v7}, 2015.

\bibitem{Censi2016}
------, ``Monotone co-design problems; or, everything is the same,'' in
  \emph{{American Control Conference}}, 2016.

\bibitem{Censi2017b}
------, ``A class of co-design problems with cyclic constraints and their
  solution,'' \emph{{IEEE Robotics and Automation Letters}}, vol.~2, pp.
  96--103, 2017.

\bibitem{SalazarLanzettiEtAl2019}
M.~Salazar, N.~Lanzetti, F.~Rossi, M.~Schiffer, and M.~Pavone, ``Intermodal
  autonomous mobility-on-demand,'' \emph{IEEE Transactions on Intelligent
  Transportation Systems}, 2019.

\bibitem{FarahaniMiandoabchiEtAl2013}
R.~Z. Farahani, E.~Miandoabchi, W.~Y. Szeto, and H.~Rashidi, ``A review of
  urban transportation network design problems,'' \emph{{European Journal of
  Operational Research}}, vol. 229, pp. 281--302, 2013.

\bibitem{GuihaireHao2008}
V.~Guihaire and J.-K. Hao, ``Transit network design and scheduling: A global
  review,'' \emph{{Transportation Research Part B: Methodological}}, vol.~42,
  pp. 1251--1273, 2008.

\bibitem{CongDeSchutterEtAl2015}
Z.~Cong, B.~De~Schutter, and R.~Babuska, ``Co-design of traffic network
  topology and control measures,'' \emph{{Transportation Research Part C:
  Emerging Technologies}}, vol.~54, pp. 56--73, 2015.

\bibitem{Arbex2015}
R.~O. Arbex and C.~B. da~Cunha, ``Efficient transit network design and
  frequencies setting multi-objective optimization by alternating objective
  genetic algorithm,'' \emph{{Transportation Research Part B: Methodological}},
  vol.~81, pp. 355--376, 2015.

\bibitem{Sun2014}
L.~Sun, J.~G. Jin, D.-H. Lee, K.~W. Axhausen, and A.~Erath, ``Demand-driven
  timetable design for metro services,'' \emph{{Transportation Research Part C:
  Emerging Technologies}}, vol.~46, pp. 284--299, 2014.

\bibitem{Su2013}
S.~Su, X.~Li, T.~Tang, and Z.~Gao, ``A subway train timetable optimization
  approach based on energy-efficient operation strategy,'' \emph{{IEEE
  Transactions on Intelligent Transportation Systems}}, vol.~14, no.~2, pp.
  883--893, 2013.

\bibitem{NarayananEtAl2020}
S.~Narayanan, E.~Chaniotakis, and C.~Antoniou, ``Shared autonomous vehicle
  services: A comprehensive review,'' \emph{{Transportation Research Part C:
  Emerging Technologies}}, vol. 111, pp. 255--293, 2020.

\bibitem{BarriosGodier2014}
J.~A. Barrios and J.~D. Godier, ``Fleet sizing for flexible carsharing systems:
  Simulation-based approach,'' \emph{{Transportation Research Record: Journal
  of the Transportation Research Board}}, vol. 2416, pp. 1--9, 2014.

\bibitem{FagnantKockelman2018}
D.~J. Fagnant and K.~M. Kockelman, ``Dynamic ride-sharing and fleet sizing for
  a system of shared autonomous vehicles in austin, texas,''
  \emph{{Transportation}}, vol.~45, no.~1, pp. 143--158, 2018.

\bibitem{Vazifeh2018}
M.~M. Vazifeh, P.~Santi, G.~Resta, S.~H. Strogatz, and C.~Ratti, ``Addressing
  the minimum fleet problem in on-demand urban mobility,'' \emph{{Nature}},
  vol. 557, no. 7706, p. 534, 2018.

\bibitem{Boesch2016}
P.~M. Boesch, F.~Ciari, and K.~W. Axhausen, ``Autonomous vehicle fleet sizes
  required to serve different levels of demand,'' \emph{{Transportation
  Research Record: Journal of the Transportation Research Board}}, vol. 2542,
  no.~1, pp. 111--119, 2016.

\bibitem{SpieserTreleavenEtAl2014}
K.~Spieser, K.~Treleaven, R.~Zhang, E.~Frazzoli, D.~Morton, and M.~Pavone,
  ``Toward a systematic approach to the design and evaluation of automated
  mobility-on-demand systems: A case study in singapore,'' in \emph{Road
  vehicle automation}, 2014, pp. 229--245.

\bibitem{ZhangSheppardEtAl2018}
H.~Zhang, C.~J.~R. Sheppard, T.~Lipman, and S.~Moura, ``Joint fleet sizing and
  charging system planning for autonomous electric vehicles,'' \emph{{IEEE
  Transactions on Intelligent Transportation Systems}}, 2019.

\bibitem{BeaujonTurnquist1991}
G.~J. Beaujon and M.~A. Turnquist, ``A model for fleet sizing and vehicle
  allocation,'' \emph{{Transportation Science}}, vol.~25, no.~1, pp. 19--45,
  1991.

\bibitem{Wallar2019}
A.~Wallar, W.~Schwarting, J.~Alonso-Mora, and D.~Rus, ``Optimizing multi-class
  fleet compositions for shared mobility-as-a-service,'' in \emph{{Proc.\ IEEE
  Int.\ Conf.\ on Intelligent Transportation Systems}}.\hskip 1em plus 0.5em
  minus 0.4em\relax IEEE, 2019, pp. 2998--3005.

\bibitem{PintoHylandEtAl2019}
H.~K. R.~F. Pinto, M.~F. Hyland, H.~S. Mahmassani, and I.~O. Verbas, ``Joint
  design of multimodal transit networks and shared autonomous mobility
  fleets,'' \emph{{Transportation Research Part C: Emerging Technologies}},
  2019.

\bibitem{Daganzo2008}
C.~F. Daganzo and N.~Geroliminis, ``An analytical approximation for the
  macroscopic fundamental diagram of urban traffic,'' \emph{{Transportation
  Research Part B: Methodological}}, vol.~42, no.~9, pp. 771--781, 2008.

\bibitem{PIM2012}
PIM. (2012) Metrorail ridership by origin and destination. Plan It Metro. Plan
  It Metro.

\bibitem{Mas-ColellWhinstonEtAl1995}
A.~Mas-Colell, M.~D. Whinston, and J.~R. Green, \emph{Microeconomic
  Theory}.\hskip 1em plus 0.5em minus 0.4em\relax {Oxford Univ.\ Press}, 1995.

\bibitem{Richards2010}
D.~C. Richards, ``Relationship between speed and risk of fatal injury:
  Pedestrians and car occupants,'' Department for Transport: London, Tech.
  Rep., 2010.

\bibitem{Ruch2018}
C.~Ruch, S.~H{\"o}rl, and E.~Frazzoli, ``Amodeus, a simulation-based testbed
  for autonomous mobility-on-demand systems,'' in \emph{{Proc.\ IEEE Int.\
  Conf.\ on Intelligent Transportation Systems}}, 2018, pp. 3639--3644.

\bibitem{HaklayWeber2008}
M.~Haklay and P.~Weber, ``{OpenStreetMap}: User-generated street maps,''
  \emph{{IEEE Pervasive Computing}}, vol.~7, no.~4, pp. 12--18, 2008.

\bibitem{GTFS2019}
GTFS. (2019) Gtfs: Making public transit data universally accessible.

\bibitem{ODDC2017}
ODDC. (2017) Taxicab trips in 2016. Open Data DC. Open Data DC. Available
  online at \url{https://opendata.dc.gov/search?q=taxicabs}.

\bibitem{Siddiqui2018b}
F.~Siddiqui. (2018) As ride hailing booms in d.c., it's not just eating in the
  taxi market -- it's increasing vehicle trips. The Washington Post. The
  Washington Post. {available online}.

\bibitem{DoA1977}
DoA, Ed., \emph{Military Police Traffic Operations}.\hskip 1em plus 0.5em minus
  0.4em\relax Department of the Army, 1977.

\bibitem{DixonIrshadEtAl2018}
S.~Dixon, H.~Irshad, and V.~White, ``Deloitte city moblity index -- washington
  d.c.'' Deloitte, Tech. Rep., 2018.

\bibitem{Censi2019}
A.~Censi. (2019) Monotone co-design problems. Available online:
  \url{https://co-design.science/index.html}.

\bibitem{PavlenkoSlowikEtAl2019}
N.~Pavlenko, P.~Slowik, and N.~Lutsey, ``When does electrifying shared mobility
  make economic sense?'' The International Council on Clean Transportation,
  Tech. Rep., 2019.

\bibitem{BoeschBeckerEtAl2018}
P.~M. Boesch, F.~Becker, H.~Becker, and K.~W. Axhausen, ``Cost-based analysis
  of autonomous mobility services,'' \emph{{Transport Policy}}, vol.~64, pp.
  76--91, 2018.

\bibitem{FagnantKockelman2015}
D.~J. Fagnant and K.~Kockelman, ``Preparing a nation for autonomous vehicles:
  opportunities, barriers and policy recommendations,'' \emph{{Transportation
  Research Part A: Policy and Practice}}, vol.~77, pp. 167--181, 2015.

\bibitem{BauerGreenblattEtAl2018}
G.~S. Bauer, J.~B. Greenblatt, and B.~F. Gerke, ``Cost, energy, and
  environmental impact of automated electric taxi fleets in manhattan,''
  \emph{{Environmental Science \& Technology}}, vol.~52, no.~8, pp. 4920--4928,
  2018.

\bibitem{Wadud2017}
Z.~Wadud, ``Fully automated vehicles: A cost of ownership analysis to inform
  early adoption,'' \emph{{Transportation Research Part A: Policy and
  Practice}}, vol. 101, pp. 163--176, 2017.

\bibitem{Litman2019}
T.~Litman, ``Autonomous vehicle implementation predictions -- implications for
  transport planning,'' Victoria Transport Policy Institute, Tech. Rep., 2019.

\bibitem{Watttime2018}
W.~Time. (2018, Mar.) Carbon footprint data. Wired.

\bibitem{WMATA2017}
WMATA, ``Fy2018 proposed budget,'' Washington Metropolitan Area Transit
  Authority, Tech. Rep., 2017.

\bibitem{Aratani2015}
L.~Aratani. (2015) Metro to debut first of its 7000-series cars on blue line on
  april 14. The Washington Post. {available online}.

\bibitem{WMATA2018}
WMATA, ``Sustainability report 2018,'' Washington Metropolitan Area Transit
  Authority, Tech. Rep., 2018.

\bibitem{Jaffe2015}
E.~Jaffe. (2015) The case for driverless trains, by the numbers. Citylab.
  Citylab. Available online.

\bibitem{HowardSylvan2015}
P.~Howard and D.~Sylvan, ``Expert consensus on the economics of climate
  change,'' Institute for Policy Integrity -- New York University School of
  Law, Tech. Rep., 2015.

\bibitem{WangZhangEtAl2016}
Y.~Wang, J.~Zhang, M.~Ma, and X.~Zhoum, ``Survey on driverless train operation
  for urban rail transit systems,'' \emph{Urban Rail Transit}, vol.~2, no.
  3--4, p. 106–113, 2016.

\bibitem{GawronKeoleianEtAl2018}
J.~H. Gawron, G.~A. Keoleian, R.~D. De~Kleine, T.~J. Wallington, and
  K.~Hyung~Chul, ``Life cycle assessment of connected and automated vehicles:
  Sensing and computing subsystem and vehicle level effects,''
  \emph{{Environmental Science \& Technology}}, vol.~52, pp. 3249--3256, 2018.

\bibitem{Lienert2019}
P.~Lienert. (2019) Cost of driverless vehicles to drop dramatically: Delphi
  ceo. Insurance Journal. {available online}.

\bibitem{WCP2018}
WCP, ``The automotive lidar market,'' Woodside Capital Partners, Tech. Rep.,
  2018.

\end{thebibliography}

\end{document}